\documentclass{article}

\usepackage{amssymb,amsfonts,amsmath}
\usepackage{cite,enumerate,float,indentfirst}
\usepackage{color}

\def\be{\begin{eqnarray}}
\def\ee{\end{eqnarray}}
\def\nn{\nonumber}

\def\Tr{{\rm Tr}\,}

\def\Sch{{\rm Schur}}

\def\beq{\be }
\def\eeq{\ee}
\def\beqa{\be }
\def\eeqa{\ee}
\def\CR{\nn\\}

\definecolor{red}{rgb}{1,0,0}
\definecolor{orange}{rgb}{1,0.5,0}
\definecolor{violet}{rgb}{0.7,0,1}

\hoffset= - 1.0in         

\begin{document}

\title{
\Large{ \bf A non-torus link from topological vertex
}}

\author{
{\bf Hidetoshi Awata$^a$}\footnote{awata@math.nagoya-u.ac.jp},
\ {\bf Hiroaki Kanno$^{a,b}$}\footnote{kanno@math.nagoya-u.ac.jp},
\ {\bf Andrei Mironov$^{c,d,e}$}\footnote{mironov@lpi.ru; mironov@itep.ru},\\
\ {\bf Alexei Morozov$^{d,e}$}\thanks{morozov@itep.ru},
\ \ and \ {\bf Andrey Morozov$^{d,e,f}$}\footnote{andrey.morozov@itep.ru}
\date{ }
}

\maketitle

\vspace{-5.2cm}

\begin{center}
\hfill FIAN/TD-08/18\\
\hfill IITP/TH-10/18\\
\hfill ITEP/TH-13/18
\end{center}

\vspace{4cm}

\begin{center}
$^a$ {\small {\it Graduate School of Mathematics, Nagoya University,
Nagoya, 464-8602, Japan}}\\
$^b$ {\small {\it KMI, Nagoya University,
Nagoya, 464-8602, Japan}}\\
$^c$ {\small {\it Lebedev Physics Institute, Moscow 119991, Russia}}\\
$^d$ {\small {\it ITEP, Moscow 117218, Russia}}\\
$^e$ {\small {\it Institute for Information Transmission Problems, Moscow 127994, Russia}}\\
$^f$ {\small {\it MIPT, Dolgoprudny, 141701, Russia}}
\end{center}

\vspace{.5cm}

\begin{abstract}
The recently suggested tangle calculus for knot polynomials
is intimately related to topological string considerations and can help to build the HOMFLY-PT invariants from the topological vertices.
We discuss this interplay in the simplest example of the Hopf link
and link $L_{8n8}$. It turns out that the resolved conifold with four different representations on
the four external legs, on the topological string side, is described by a special projection of the four-component link $L_{8n8}$, which reduces to the Hopf link colored with two composite representations.
Thus, this provides the first {\it explicit} example of non-torus link description through topological vertex. It is not a real breakthrough, because $L_{8n8}$ is just a cable of the Hopf link, still it can help to intensify the development of  the formalism towards more interesting examples.
\end{abstract}

\section{Introduction}

Correlation functions of Wilson loops are the most interesting observables
in gauge theories: these are the gauge invariant quantities needed to understand confinement
and various phase transitions.
Unfortunately, they are
much more difficult to calculate, even approximately.
Therefore, of interest are particular cases, where Wilson loop averages are exactly
calculable, and one can hope to develop a formalism capturing and efficiently
describing the peculiar properties of these non-local observables.
The best known example of such exactly solvable problem is $3d$ Chern-Simons theory
\cite{CS}, which is topological and essentially Gaussian in particular gauges like ${\cal A}_0=0$,
which allows one to formulate it in very different dual terms,
for instance, as the Reshetikhin-Turaev (RT) lattice theory  \cite{RT} on 4-valent graphs,
which are the knot/link diagrams in projection from three to two spatial dimensions.
In result, the correlators in simply-connected $3d$ space-time are essentially rational
functions with well controlled denominators, naturally called knot/link polynomials
\cite{knotpols,Con}, being investigated in knot theory for already about a century.

One of spectacular results of the RT approach in its modern Tanaka-Krein version \cite{RTmod}
is a possibility of defining and exploiting ``observables", arising when the
Wilson loops are cut in pieces, into Wilson lines.
An advantage is that one can then construct the original correlators by "gluing" them from
much simpler building blocks.
The problem is, however, that, in gauge theory, the open loops provide gauge-non-invariant quantities
which can not be ascribed any meaning, neither physical, nor mathematical.
However, a reformulation in RT terms, where everything is automatically gauge invariant,
allows one to bypass this difficulty and introduce ``tangle blocks" \cite{Con,ML,tangcalc1},
which has no clear definition in the original Chern-Simons theory, however, can be
efficiently used in cut-and-glue procedures for constructing and evaluating link invariants \cite{ML,tangcalc}.
A great capability of this approach was already demonstrated by developing
the arborescent calculus \cite{arbor}, which is currently the most advanced
working method to evaluate colored link invariants.
The main task of the tangle calculus is, however, more ambitious:
it should help to understand a complicated network of non-linear
relations between various knot invariants for different links/knots and
different representations, and provide a closed self-reliable theory of
Wilson loop correlators, which does not explicitly refer to the ``mother"
Chern-Simons model, relevant for a perturbative description of one particular phase
of the entire theory.
Unfortunately, the method is still limited by the lack of general view
and of clear relation to other approaches.

The goal of the present letter is to describe such relations, and even, in the simplest possible examples, this provides new insights and
new results. In particular, we discuss a relation to another framework: the framework of
topological strings, dual to Chern-Simons theory \cite{GV,OV}.  The main object in this theory is the topological vertex \cite{tv,Aganagic:2003db} (see \cite{IKV,GIKV,AK0,AK} for its refined version),  and \cite{Zenk} for related network models, which are arbitrary convolutions of vertices. It is a long-standing problem to express the knot and link invariants in these terms, and, more generally, in terms of arbitrary tangle blocks. The simplest example is provided by the resolved conifold constructed from just a pair of such vertices. It is well known \cite{tv,Aganagic:2003db} that, for a particular choice of two non-trivial representations on external legs, the answer coincides with that for the 2-component Hopf link, while for arbitrary choice one could rather expect the 4-component link $L_{8n8}$ . See also \cite{BFM}, where the resolved conifold with four non-trivial representations was considered as a covering contribution to the Hopf link of $SO/Sp$ Chern-Simons theory. We demonstrate in this paper that the actual  relation is less trivial: the resolved conifold stays related to the Hopf link, only composite representations get involved. As to $L_{8n8}$, only a piece of its invariant is reproduced in this way, while the entire expression is rather a sum of reduced conifold contributions over a set of representations at external legs. In fact, $L_{8n8}$ is a cable of the Hopf link, thus it is not a big surprise that these invariants are related this way, still this is the first example of a formula for a non-torus knot/link constructed from topological vertices.

Thus, the main claim of the paper is that the resolved conifold with branes on the four external legs $Z_{\mu_1,\mu_2;\lambda_1,\lambda_2}$, which is a function of four representations $\lambda_{1,2}$, $\mu_{1,2}$, describes a special (in a sense, maximal) projection of the colored HOMFLY invariant ${\cal H}^{L_{8n8}}$ of the link $L_{8n8}$ (in accordance with the Thistlethwaite link table \cite{twi}) in the colored space, which ultimately reduces to the HOMFLY invariant of the Hopf link ${\cal H}^{\rm Hopf}$ with the two components colored by the composite representations $(\lambda_1,\lambda_2)$ and $(\mu_1,\mu_2)$:
\be
\boxed{\boxed{
{\cal G}^{L_{8n8}}_{\mu_1\times\lambda_1\times\mu_2\times\lambda_2}: =
\hbox{\bf Pr}\left[{\cal H}^{L_{8n8}}_{\mu_1\times\lambda_1\times\mu_2\times\lambda_2}\right]_{\rm max}=
\frac{Z_{\mu_1,\mu_2;\lambda_1,\lambda_2}}
{Z_{\varnothing,\varnothing:\varnothing, \varnothing}} ={\cal H}^{\rm Hopf}_{(\lambda_1,\lambda_2)\times(\mu_1,\mu_2)}
}}
\ee
Note that this is an open question if there is a possibility of expressing generic knots and links through topological vertices within the network model framework. The only examples known so far are the torus knots/links, and their invariants are either constructed from the Chern-Simons $S$-matrix \cite{tv,Aganagic:2003db,AS} which is the Hopf invariant and is associated with the topological vertex only in this way, or constructed \cite{Klemm} by basically substituting the Adams coefficients into the Rosso-Jones formula \cite{RJ}, with the second ingredient being eigenvalues of the cut-and-join (or Casimir) operator, which do not have any explicit meaning in topological calculus, and are just introduced "by hands". Hence, even in the case of torus knots and links it remains unclear if they can be obtained directly via network configurations. Specifics of the Hopf link invariant is that, in this case, the Rosso-Jones formula, involving a Casimir-weighted sum  of the Schur functions in various representations, accidentally (?) coincides with just a single Schur function at a peculiarly deformed topological locus, and thus acquires a simple representation theory meaning.

The plan of the paper is as follows: in sections 2 and 3, we describe the HOMFLY Hopf invariant. Then, in section 4, the Hopf link is related with the conifold description, which naturally gives rise to link $L_{8n8}$ in the general case. In fact, it turns out that, from the conifold approach, one obtains not the full colored HOMFLY invariant for the $L_{8n8}$ link, but its projection, which is ultimately projection to composite representation. Hence, one needs to consider the Hopf link colored with composite representations. It is done in section 5. After this preliminary work, in section 6, we are able to formulate our main statement: that the conifold consideration through the topological vertex, indeed, gives rise to the projected $L_{8n8}$ link aka the Hopf link in composite representations. This statement is proved in sections 7-8. Section 9 described symmetry properties of the HOMFLY Hopf invariant. At last, in section 10, we discuss relations of the description HOMFLY Hopf invariant as the Schur function and the Rosso-Jones formula \cite{RJ} and generalization to the superpolynomials.
The Appendix contains illustrative examples of the main statement of this paper for first representations.

\section{Hopf recursion}

The simplest relation provided by the tangle calculus of \cite{tangcalc}
arises when one connects the open ends of the Hopf tangle

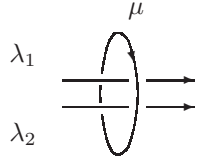
\begin{picture}(400,110)(-200,-60)

\put(0,0){
\put(0,5){\line(1,0){25}}
\put(32,5){\vector(1,0){18}}
\put(25,-5){\line(-1,0){25}}
\put(32,-5){\vector(1,0){18}}

\qbezier(25,-20)(32,0)(25,20)
\qbezier(25,-20)(18,-30)(15,-8)
\qbezier(25,20)(18,30)(15,8)
\qbezier(14.5,-3)(14.2,0)(14.5,3)

\put(-20,12){\mbox{$\lambda_1$}}
\put(-20,-18){\mbox{$\lambda_2$}}
\put(25,30){\mbox{$\mu$}}
\put(25,20){\vector(1,-4){2}}
}

\end{picture}

\noindent
and treats it in two different ways:

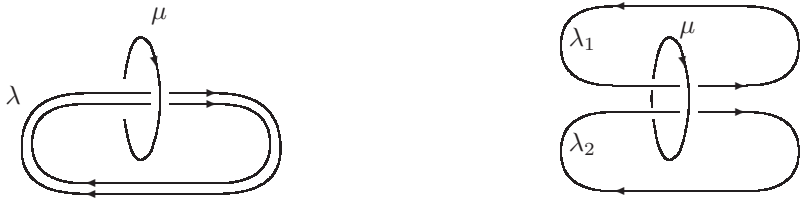
\begin{picture}(400,110)(-200,-60)

\put(-100,0){
\put(0,2){\line(1,0){25}}
\put(32,2){\vector(1,0){18}}
\put(25,-2){\line(-1,0){25}}
\put(32,-2){\vector(1,0){18}}

\qbezier(25,-20)(32,0)(25,20)
\qbezier(25,-20)(18,-30)(15,-8)
\qbezier(25,20)(18,30)(15,8)

\put(0,3){
\qbezier(0,-5)(-20,-5)(-20,-20)
\qbezier(0,-35)(-20,-35)(-20,-20)
\put(50,-35){\vector(-1,0){50}}
\qbezier(50,-5)(70,-5)(70,-20)
\qbezier(50,-35)(70,-35)(70,-20)
}

\qbezier(0,2)(-24,2)(-24,-19)
\qbezier(0,-36)(-24,-36)(-24,-19)
\put(50,-36){\vector(-1,0){50}}
\qbezier(50,2)(74,2)(74,-19)
\qbezier(50,-36)(74,-36)(74,-19)

\put(-30,-2){\mbox{$\lambda$}}
\put(25,30){\mbox{$\mu$}}
\put(25,20){\vector(1,-4){2}}

}

\put(100,0){
\put(0,5){\line(1,0){25}}
\put(32,5){\vector(1,0){18}}
\put(25,-5){\line(-1,0){25}}
\put(32,-5){\vector(1,0){18}}

\qbezier(25,-20)(32,0)(25,20)
\qbezier(25,-20)(18,-30)(15,-8)
\qbezier(25,20)(18,30)(15,8)
\qbezier(14.5,-3)(14.2,0)(14.5,3)

\qbezier(0,-5)(-20,-5)(-20,-20)
\qbezier(0,-35)(-20,-35)(-20,-20)
\put(50,-35){\vector(-1,0){50}}
\qbezier(50,-5)(70,-5)(70,-20)
\qbezier(50,-35)(70,-35)(70,-20)

\qbezier(0,5)(-20,5)(-20,20)
\qbezier(0,35)(-20,35)(-20,20)
\put(50,35){\vector(-1,0){50}}
\qbezier(50,5)(70,5)(70,20)
\qbezier(50,35)(70,35)(70,20)

\put(-17,20){\mbox{$\lambda_1$}}
\put(-17,-20){\mbox{$\lambda_2$}}
\put(25,25){\mbox{$\mu$}}
\put(25,20){\vector(1,-4){2}}

}

\end{picture}

\noindent
The 3-component link at the right is composite, i.e. its {\it reduced} HOMFLY ``polynomial"\footnote{The word ``polynomial" should not lead to a confusion here: the colored HOMFLY invariants are, in fact, rational functions of $q$ and $A=q^N$ with simple denominators proportional to quantum dimensions. Hereafter, we call them HOMFLY polynomials in order to use the same term both for links and knots (when the reduced HOMFLY invariant is, indeed, a Laurent polynomial of $A$ and $q$).}
is just a product of two, for the two constituent Hopf links.
At the left, we treat the same configuration as a single Hopf link,
where reducible is the representation $\lambda_1\otimes\lambda_2$,
which can be decomposed into irreps, thus the HOMFLY polynomial is a sum
of Hopf polynomials in different representations, taken with appropriate
multiplicities $N^\lambda_{\lambda_1\lambda_2}$ (which actually are all unities
if either $\lambda_1$ or $\lambda_2$ is symmetric, or the both are rectangular).
Therefore, we get a relation between the {\it unreduced} Hopf polynomials
\be
\boxed{
D_\mu\cdot \!\!\!\!
\sum_{\lambda\in \lambda_1\otimes\lambda_2}N^\lambda_{\lambda_1\lambda_2}\cdot
{\cal H}^{\rm Hopf}_{\lambda,\mu}
\ \ = \ \ {\cal H}^{\rm Hopf}_{\lambda_1,\mu}\cdot{\cal H}^{\rm Hopf}_{\lambda_2,\mu}
}
\label{HvsHH}
\ee
where $D_\mu$ is the quantum dimension of the representation $\mu$.
This relation is correct for the Hopf polynomial in the standard, or canonical framing \cite{Atiah,MarF,China1,tangcalc}, which we use throughout the paper.

It can be used as a recursion providing efficient and explicit
formulas for colored Hopf invariants \cite{tangcalc},
which do not involve any functions difficult to deal with,  like the Schur polynomials.
Its $t$-deformation changes the coefficients in the sum over $\lambda$,
still, it seems to persist for colored superpolynomials
available from \cite{DMMSS,Ch,MMSS,GN}.
This is non-trivial, because there is no reason why the Khovanov-Rozansky
cohomological calculus, underlying today's theory of superpolynomials,
should respect cut-and-gluing procedures of the effective lattice theory
that are behind the Reshetikhin-Turaev formalism.
Indeed, it is not {\it quite} respected, because the coefficients
are deformed, but still it survives, since non-linear relations
like (\ref{HvsHH}) continue to exist.

\section{Recursion from character decomposition}

On the other hand, the topological vertex formalism \cite{tv,Aganagic:2003db,IKV,GIKV,AK,IK} is associated with an
expression for the Hopf polynomials through characters, Schur or Macdonald
functions, which is not very efficient in description of knot polynomials
but is very convenient for gluing
via the Cauchy summation formulas.
Namely, according to \cite{ML,Mar,GIKV,AK},
\be
{\cal H}_{\lambda,\mu}^{\rm Hopf} =q^{2|\lambda||\mu|\over N}\cdot
\Sch_\lambda(q^{-\rho})\cdot \Sch_\mu(q^{-\lambda-\rho})
\label{HopfthroughSchur}
\ee
where $\rho$ is the Weyl vector (half-sum of all positive roots) and $\lambda$, $\mu$ are the weights of representations, and $\Sch_R(x_i)$ is the Schur function associated with the Young diagram $R$, i.e. the character of  the $U(N)$ group representation associated with $R$ \cite{Sch}. $\Sch_R(x_i)$ is here a symmetric function of $N$ variables $x_i^{2}$, $i=1...N$ associated with components of the corresponding vectors in the Cartan plane. For instance, $2\rho=(N-1,N-3,\ldots,-N+3,-N+1)=\{N-2i+1\}$. In (\ref{HopfthroughSchur}), we manifestly have taken into account the $U(1)$-factor $q^{2|\lambda||\mu|\over N}$, \cite{Atiah,MarF,China1,tangcalc}.

The Hopf polynomial (\ref{HopfthroughSchur}) depends on variables $q$ and $A=q^N$. The drawback of this presentation is that (\ref{HopfthroughSchur}) is defined for a concrete $N$, and obtaining its $A$-dependence requires an analytic continuation. Instead, one can proceed with a presentation with the number of variables in the Schur function not related with {\it the parameter} $A$, \cite{AK}. To this end, one has to introduce the shifted Weyl vector $2\rho_0=(-1,-3,\ldots)=\{-2i+1\}$, and consider the Schur functions as symmetric functions of two sets of variables:
\be
{\cal H}_{\lambda,\mu}^{\rm Hopf} =q^{2|\lambda||\mu|\over N}\cdot
\Sch_\lambda(A^{-1}q^{-\rho_0},Aq^{\rho_0})\cdot \Sch_\mu(A^{-1}q^{-\lambda-\rho_0},Aq^{\rho_0})
\ee
and now one may not restrict by $N$ the number of variables in the Schur symmetric function.

The Hopf polynomial is, in fact, symmetric under the permutation of $\lambda$ and $\mu$, i.e.
\be
\Sch_\lambda(q^{-\rho})\cdot \Sch_\mu(q^{-\lambda-\rho})
= \Sch_\mu(q^{-\rho})\cdot \Sch_\lambda(q^{-\mu-\rho}).
\label{symrel}
\ee
Then
\be
{\cal H}_{\lambda_1,\mu}^{\rm Hopf} \cdot {\cal H}_{\lambda_2,\mu}^{\rm Hopf} =q^{2(|\lambda_1|+|\lambda_2|)|\mu|\over N}\cdot
\Sch_\mu(q^{-\rho})
\cdot \Sch_\mu(q^{-\rho})\cdot
\Sch_{\lambda_1}(q^{-\mu-\rho})\cdot
\Sch_{\lambda_2}(q^{-\mu-\rho}).
\ee
Since we made a clever choice between the two versions in (\ref{symrel}),
the arguments of $\Sch_{\lambda_1}$ and $\Sch_{\lambda_2}$
depend on the same diagram $\mu$, i.e. coincide,
and one can use the multiplication rule
\be
\Sch_{\lambda_1}\cdot \Sch_{\lambda_2}
= \sum_{\lambda\in \lambda_1\otimes\lambda_2}
N^\lambda_{\lambda_1\lambda_2}\cdot\Sch_\lambda
\ee
where $N^\lambda_{\lambda_1\lambda_2}$ are integer-valued Littlewood-Richardson coefficients.
This gives:
\be
{\cal H}_{\lambda_1,\mu}^{\rm Hopf} \cdot {\cal H}_{\lambda_2,\mu}^{\rm Hopf}
&&=
\sum_{\lambda\in \lambda_1\otimes\lambda_2}
N^\lambda_{\lambda_1\lambda_2}\cdot
\overbrace{\Sch_\mu(q^{ -\rho})}^{\cdot D_\mu}\cdot
\overbrace{q^{2(|\lambda_1|+|\lambda_2|)|\mu|\over N}\cdot\Sch_\mu(q^{ -\rho})\cdot \Sch_\lambda(q^{-\mu-\rho})
}^{
{\cal H}_{\lambda,\mu}^{\rm Hopf}} \CR
&&=D_\mu \cdot\!\!\!\!
\sum_{\lambda\in \lambda_1\otimes\lambda_2}
N^\lambda_{\lambda_1\lambda_2} \cdot {\cal H}_{\lambda,\mu}^{\rm Hopf}
\label{HvsHH2}
\ee
i.e. reproduces (\ref{HvsHH}).

One can also use the time variables $p_k:=\sum_i x_i^{2k}$ so that the quantum dimension is
\be
D_\mu
=\Sch_\mu\{p^*\},\ \ \ \ \ \ \ \ p_k^*:= \frac{A^k-A^{-k}}{q^k-q^{-k}}
\ee
Similarly, one can consider the time variables for $q^{-\lambda-\rho}$, moreover, one can absorb the $U(1)$-factor (\ref{HopfthroughSchur}) in their definition so that
\be
{p_k^{*\lambda}} = q^{2|\lambda|k\over N}\Big(p^*_k - A^{-k}(q^k-q^{-k})\sum_{i,j\in\lambda}q^{2k(i-j)}\Big)=
q^{2|\lambda|k\over N}\Big(p^*_k + A^{-k}\sum_i q^{(2i-1)k}(q^{-2k\lambda_i}-1)\Big)
\label{plambdatimes}
\ee
where $\lambda_i$ are lengths of lines of the Young diagram $\lambda$, $|\lambda|=\sum_i\lambda_i$ and $A=q^N$.
In terms of these time variables
\be\label{10}
{\cal H}^{\rm Hopf}_{\mu\lambda}=D_\lambda \cdot \Sch_\mu\{p^{*\lambda}\}
\ee
For instance,
\be
\frac{{\cal H}_{[1],[1]}^{\rm Hopf}}{D_{[1]}} = p_1^{*[1]} =q^{2\over N}\
\frac{A-A^{-1}(q^2-1+q^{-2})}{q-q^{-1}}
\ee
If $\lambda$ is the adjoint representation, then the {\it uniform} Hopf polynomial
\cite{MMkrM} is associated with the time variables
\be
p^{*{\rm adj}}_k = (q^{2k}-1+q^{-2k})p_k^*
\ee
see \cite{tangcalc,MMHopf} for further details.

\section{Conifold description
\label{coni}}

In topological vertex theory of \cite{tv,Aganagic:2003db,IKV,GIKV,AK}, the Hopf polynomial
${\cal H}_{\lambda,\mu}^{\rm Hopf}$ is associated
with the brane pattern, symbolically depicted as

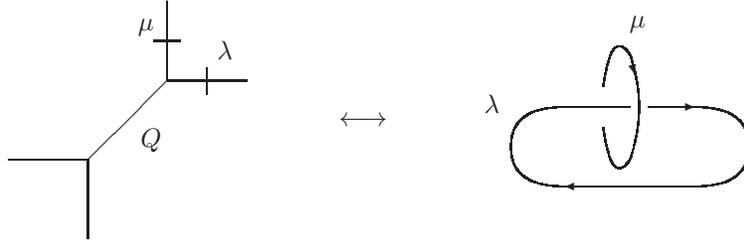
\begin{picture}(300,100)(-150,-70)

\put(0,0){\line(1,0){30}}
\put(0,0){\line(0,1){30}}
\put(0,0){\line(-1,-1){30}}
\put(-30,-30){\line(-1,0){30}}
\put(-30,-30){\line(0,-1){30}}
\put(-10,-25){\mbox{$Q $}}

\put(-5,15){\line(1,0){10}} \put(-11,19){\mbox{$\mu$}}
\put(15,-5){\line(0,1){10}} \put(19,8){\mbox{$\lambda$}}

\put(65,-17){\mbox{$\longleftrightarrow$}}

\put(150,-10){
\put(0,0){\line(1,0){25}}
\put(32,0){\vector(1,0){18}}

\qbezier(25,-20)(32,0)(25,20)
\qbezier(25,-20)(18,-30)(15,-8)
\qbezier(25,20)(18,30)(15,8)

\put(0,5){
\qbezier(0,-5)(-20,-5)(-20,-20)
\qbezier(0,-35)(-20,-35)(-20,-20)
\put(50,-35){\vector(-1,0){50}}
\qbezier(50,-5)(70,-5)(70,-20)
\qbezier(50,-35)(70,-35)(70,-20)
}

\put(-30,-2){\mbox{$\lambda$}}
\put(25,30){\mbox{$\mu$}}
\put(25,20){\vector(1,-4){2}}

}

\end{picture}

\noindent
According to our logic in the present paper, $\lambda$ can actually be a representation
from the product of two.
In fact, the same can be true for $\mu$.
In other words, we can consider four  branes put on the external legs:

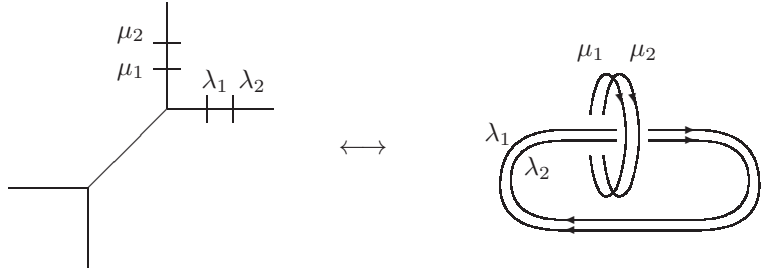
\begin{picture}(300,120)(-150,-70)

\put(0,0){\line(1,0){40}}
\put(0,0){\line(0,1){40}}
\put(0,0){\line(-1,-1){30}}
\put(-30,-30){\line(-1,0){30}}
\put(-30,-30){\line(0,-1){30}}

\put(-5,15){\line(1,0){10}} \put(-19,13){\mbox{$\mu_1$}}
\put(15,-5){\line(0,1){10}} \put(13,8){\mbox{$\lambda_1$}}

\put(-5,25){\line(1,0){10}} \put(-19,27){\mbox{$\mu_2$}}
\put(25,-5){\line(0,1){10}} \put(27,8){\mbox{$\lambda_2$}}

\put(65,-17){\mbox{$\longleftrightarrow$}}

\put(150,-10){
\put(0,2){\line(1,0){20}}
\put(32,2){\vector(1,0){18}}
\put(20,-2){\line(-1,0){20}}
\put(32,-2){\vector(1,0){18}}

\qbezier(20,-20)(27,0)(20,20)
\qbezier(20,-20)(13,-30)(10,-8)
\qbezier(20,20)(13,30)(10,8)

\qbezier(25,-20)(32,0)(25,20)
\qbezier(25,-20)(18,-30)(15,-8)
\qbezier(25,20)(18,30)(15,8)

\put(0,3){
\qbezier(0,-5)(-20,-5)(-20,-20)
\qbezier(0,-35)(-20,-35)(-20,-20)
\put(50,-35){\vector(-1,0){50}}
\qbezier(50,-5)(70,-5)(70,-20)
\qbezier(50,-35)(70,-35)(70,-20)
}

\qbezier(0,2)(-24,2)(-24,-19)
\qbezier(0,-36)(-24,-36)(-24,-19)
\put(50,-36){\vector(-1,0){50}}
\qbezier(50,2)(74,2)(74,-19)
\qbezier(50,-36)(74,-36)(74,-19)

\put(-30,-2){\mbox{$\lambda_1$}} \put(-15,-15){\mbox{$\lambda_2$}}
\put(5,30){\mbox{$\mu_1$}}\put(25,30){\mbox{$\mu_2$}}
\put(25,20){\vector(1,-4){2}}
\put(20,20){\vector(1,-4){2}}

}

\end{picture}

\noindent
and interpret this picture as the double sum

\begin{picture}(300,130)(-240,-80)

\put(-250,-17){\mbox{$\sum_{\lambda\in\lambda_1\otimes \lambda_2}
\sum_{\mu\in \mu_1\otimes\mu_2} \ \
N^{\lambda}_{\lambda_1\lambda_2}\cdot N^\mu_{\mu_1\mu_2}\cdot
$}}

\qbezier(-70,-70)(-90,-15)(-70,40)
\qbezier(200,-70)(220,-15)(200,40)

\put(0,0){\line(1,0){30}}
\put(0,0){\line(0,1){30}}
\put(0,0){\line(-1,-1){30}}
\put(-30,-30){\line(-1,0){30}}
\put(-30,-30){\line(0,-1){30}}

\put(-5,15){\line(1,0){10}} \put(-11,19){\mbox{$\mu$}}
\put(15,-5){\line(0,1){10}} \put(19,8){\mbox{$\lambda$}}

\put(50,-17){\mbox{$\longleftrightarrow$}}

\put(120,-10){
\put(0,0){\line(1,0){25}}
\put(32,0){\vector(1,0){18}}

\qbezier(25,-20)(32,0)(25,20)
\qbezier(25,-20)(18,-30)(15,-8)
\qbezier(25,20)(18,30)(15,8)

\put(0,5){
\qbezier(0,-5)(-20,-5)(-20,-20)
\qbezier(0,-35)(-20,-35)(-20,-20)
\put(50,-35){\vector(-1,0){50}}
\qbezier(50,-5)(70,-5)(70,-20)
\qbezier(50,-35)(70,-35)(70,-20)
}

\put(-30,-2){\mbox{$\lambda$}}
\put(25,30){\mbox{$\mu$}}
\put(25,20){\vector(1,-4){2}}

}

\end{picture}

\noindent

Alternatively, one can drag two of these four branes through to the other external legs in order to get:

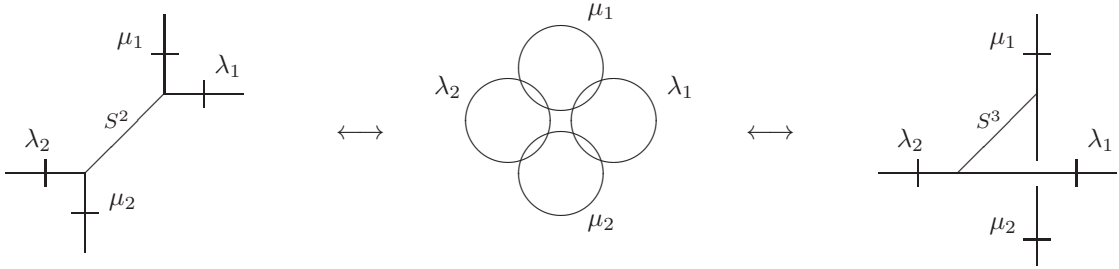
\begin{picture}(300,120)(-90,-75)

\put(0,0){\line(1,0){30}}
\put(0,0){\line(0,1){30}}
\put(0,0){\line(-1,-1){30}}
\put(-30,-30){\line(-1,0){30}}
\put(-30,-30){\line(0,-1){30}}

\put(-5,15){\line(1,0){10}} \put(-18,19){\mbox{$\mu_1$}}
\put(15,-5){\line(0,1){10}} \put(19,8){\mbox{$\lambda_1$}}

\put(-35,-45){\line(1,0){10}} \put(-21,-42){\mbox{$\mu_2$}}
\put(-45,-35){\line(0,1){10}} \put(-53,-20){\mbox{$\lambda_2$}}

\put(-23,-13){\footnotesize\mbox{$S^2$}}

\put(65,-17){\mbox{$\longleftrightarrow$}}

\put(150,-10){

\put(0,20){\circle{30}}
\put(-20,0){\circle{30}}
\put(20,0){\circle{30}}
\put(0,-20){\circle{30}}

\put(10,40){\mbox{$\mu_1$}}
\put(40,10){\mbox{$\lambda_1$}}
\put(10,-40){\mbox{$\mu_2$}}
\put(-48,10){\mbox{$\lambda_2$}}

}

\put(220,-17){\mbox{$\longleftrightarrow$}}

\put(330,0){
\put(-30,-30){\line(1,0){60}}
\put(-30,-30){\line(-1,0){30}}
\put(0,0){\line(0,1){30}}
\put(0,0){\line(0,-1){25}}
\put(0,-35){\line(0,-1){30}}
\put(0,0){\line(-1,-1){30}}

\put(-5,15){\line(1,0){10}} \put(-18,19){\mbox{$\mu_1$}}
\put(15,-35){\line(0,1){10}} \put(19,-20){\mbox{$\lambda_1$}}

\put(-5,-55){\line(1,0){10}} \put(-18,-51){\mbox{$\mu_2$}}
\put(-45,-35){\line(0,1){10}} \put(-53,-20){\mbox{$\lambda_2$}}

\put(-23,-13){\footnotesize\mbox{$S^3$}}
}

\end{picture}

\noindent
The link on the r.h.s. is the 4-component $L_{8n8}$,
and it is obviously the same as the link in the second
pictures of the present section.
The identity reflecting the possibility of multiplying representations on one leg,
or, alternatively, the possibility of brane-dragging states that
\be
\boxed{
{\cal H}^{L_{8n8}}_{\mu_1,\lambda_1,\mu_2,\lambda_2} =
\sum_{\stackrel{\lambda\in \lambda_1\otimes\bar\lambda_2}{\mu\in \mu_1\otimes\bar\mu_2}} \ \
N^{\lambda}_{\lambda_1\lambda_2}\cdot N^\mu_{\mu_1\mu_2}\cdot
{\cal H}^{\rm Hopf}_{\lambda,\mu}
}
\label{L8n8}
\ee
and is immediately clear from the picture:

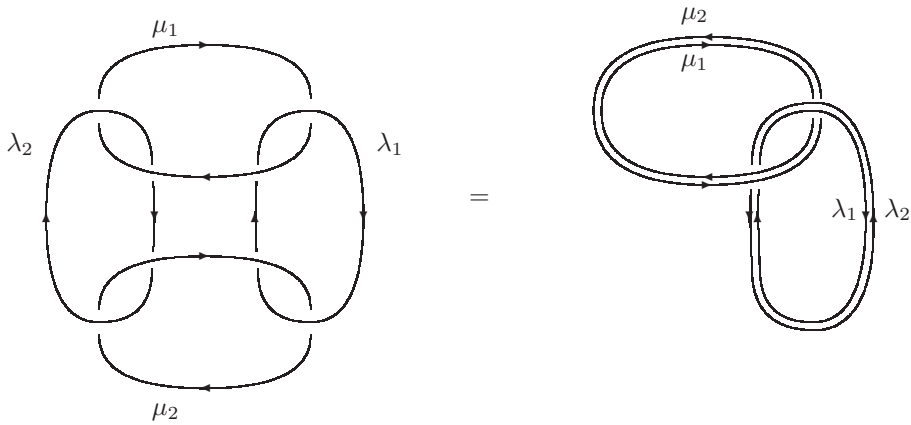
\begin{picture}(300,180)(-120,-135)

\qbezier(-40,0)(-40,20)(0,20) \qbezier(40,0)(40,20)(0,20)
\qbezier(-40,-10)(-40,-30)(0,-30) \qbezier(40,-10)(40,-30)(0,-30)

\put(0,-80){
\qbezier(-40,0)(-40,20)(0,20) \qbezier(40,0)(40,20)(0,20)
\qbezier(-40,-10)(-40,-30)(0,-30) \qbezier(40,-10)(40,-30)(0,-30)
}

\qbezier(-40,-5)(-60,-5)(-60,-45)\qbezier(-40,-85)(-60,-85)(-60,-45)
\qbezier(-40,-5)(-20,-5)(-20,-25)\qbezier(-40,-85)(-20,-85)(-20,-65)
\qbezier(-19.5,-32)(-19,-45)(-19.5,-58)

\qbezier(40,-5)(60,-5)(60,-45)\qbezier(40,-85)(60,-85)(60,-45)
\qbezier(40,-5)(20,-5)(20,-25)\qbezier(40,-85)(20,-85)(20,-65)
\qbezier(19.5,-32)(19,-45)(19.5,-58)

\put(-60,-45){\vector(0,1){2}}
\put(-19,-45){\vector(0,-1){2}}
\put(0,20){\vector(1,0){2}}
\put(0,-30){\vector(-1,0){2}}

\put(60,-45){\vector(0,-1){2}}
\put(19,-45){\vector(0,1){2}}
\put(0,-110){\vector(-1,0){2}}
\put(0,-60){\vector(1,0){2}}

\put(-75,-20){\mbox{$\lambda_2$}}
\put(-20,25){\mbox{$\mu_1$}}
\put(65,-20){\mbox{$\lambda_1$}}
\put(-20,-120){\mbox{$\mu_2$}}

\put(100,-40){\mbox{$=$}}

\put(190,0){
\qbezier(-40,-5)(-40,20)(0,20) \qbezier(40,0)(40,20)(0,20)
\qbezier(-40,-5)(-40,-30)(0,-30) \qbezier(40,-8)(40,-30)(0,-30)

\qbezier(-43,-5)(-43,23)(0,23) \qbezier(43,0)(43,23)(0,23)
\qbezier(-43,-5)(-43,-33)(0,-33) \qbezier(43,-8)(43,-33)(0,-33)

\qbezier(40,-5)(60,-5)(60,-45)\qbezier(40,-85)(60,-85)(60,-45)
\qbezier(40,-5)(20,-5)(20,-25)\qbezier(40,-85)(20,-85)(20,-65)
\qbezier(19.5,-35)(19,-45)(20,-65)

\qbezier(43,-2)(63,-5)(63,-48)\qbezier(43,-88)(63,-85)(63,-48)
\qbezier(43,-2)(17,-1)(17,-25)\qbezier(43,-88)(17,-89)(17,-65)
\qbezier(19.5,-35)(19,-45)(20,-65)
\qbezier(16.5,-35)(16,-45)(17,-65)

\put(0,20){\vector(1,0){2}}
\put(0,-30){\vector(-1,0){2}}

\put(0,23){\vector(-1,0){2}}
\put(0,-33){\vector(1,0){2}}

\put(60,-45){\vector(0,-1){2}}
\put(19,-45){\vector(0,1){2}}

\put(63,-45){\vector(0,1){2}}
\put(16,-45){\vector(0,-1){2}}

\put(47,-45){\mbox{$\lambda_1$}}
\put(-10,30){\mbox{$\mu_2$}}
\put(67,-45){\mbox{$\lambda_2$}}
\put(-10,12){\mbox{$\mu_1$}}

}

\end{picture}

\noindent
Note that the apparent cyclic symmetry of the l.h.s. in (\ref{L8n8}) is
hidden on its r.h.s.

This is an interesting formula, because one can independently calculate both sides
by the methods of \cite{tangcalc}.
The l.h.s. is a necklace with ${\cal H}^{L_{8n8}}=\Tr  \tau^4$,
where the lock-element $ \tau$,
can be extracted from the knowledge about twist knots, where
$ {\cal H}^{{\rm twist}_k}=\Tr   \tau\,\bar T^{2k}$.
The r.h.s. involves the Hopf polynomials,
which we review in the next section.
An even more interesting option is to apply conifold formulas
from \cite{GIKV,AK}, and we do this in
sec.\ref{sumfor} below.
Surprisingly or not, they do {\it not} reproduce (\ref{L8n8}),
but pick up a single item ${\cal H}^{\rm Hopf}_{(\lambda_1,\lambda_2)\times(\mu_1,\mu_2)}$
from the r.h.s. sum. We explain it in more details in the next sections.

\section{Hopf polynomials}

Colored HOMFLY polynomials for the Hopf link are in the intersection of application domains of very different approaches:
from the Rosso-Jones formula to conifold calculus.
Moreover, while the Rosso-Jones formula nicely describes ``universal" representations, i.e. the representations of $SU(N)$ that do not depend\footnote{This independence on $N$ means that the character is determined by some Young diagram(s) not involving $N$. The formal definition is related with the notion of universal character, see \cite{Koike}.} on $N$ at large enough $N$, the conifold calculus deals with the composite  (or rational, \cite{Koike,Kanno}; or coupled, \cite{Vafa}) representations, which manifestly depend on $N$, the simplest example of these being conjugate representations.

First of all, we need to explain what is the composite representation, which is, in a sense, a ``maximal" representation in the product of
$R$ and the conjugate $\bar P$ of $P$. The composite representation is the most general finite-dimensional irreducible highest weight representations of $SU(N)$  \cite{Koike,GW,Vafa,Kanno,MarK}, which are associated with the Young diagram obtained by putting $R$ atop of
$p_1$ lines of the lengths $N-p_i^{\vee}$ ($p_i^{\vee}$ are length of lines of the transposed Young diagram $P^{\vee}$), i.e.
$$(R,P)= \Big[r_1+p_1,\ldots,r_{l_R}+p_1,\underbrace{p_1,\ldots,p_1}_{N-l_{\!_R}-l_{\!_P}},
p_1-p_{_{l_{\!_P}}},p_1-p_{{l_{\!_P}-1}},\ldots,p_1-p_2\Big]$$
or, pictorially,

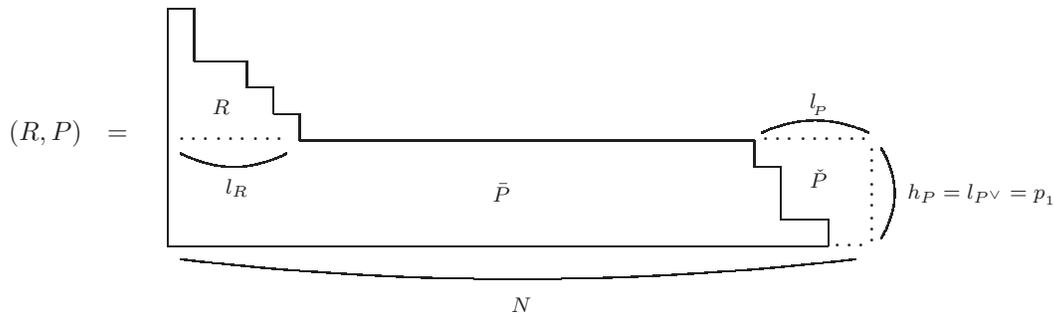
\begin{picture}(300,125)(-90,-30)

\put(0,0){\line(0,1){90}}
\put(0,0){\line(1,0){250}}
\put(50,40){\line(1,0){172}}

\put(0,90){\line(1,0){10}}
\put(10,90){\line(0,-1){20}}
\put(10,70){\line(1,0){20}}
\put(30,70){\line(0,-1){10}}
\put(30,60){\line(1,0){10}}
\put(40,60){\line(0,-1){10}}
\put(40,50){\line(1,0){10}}
\put(50,50){\line(0,-1){10}}

\put(265,2){\mbox{$\vdots$}}
\put(265,15){\mbox{$\vdots$}}
\put(265,28){\mbox{$\vdots$}}

\put(252,0){\mbox{$\ldots$}}
\put(253,40){\mbox{$\ldots$}}
\put(239,40){\mbox{$\ldots$}}
\put(225,40){\mbox{$\ldots$}}

\put(222,40){\line(0,-1){10}}
\put(222,30){\line(1,0){10}}
\put(232,30){\line(0,-1){20}}
\put(232,10){\line(1,0){18}}
\put(250,0){\line(0,1){10}}

\put(0,90){\line(1,0){10}}
\put(10,90){\line(0,-1){20}}
\put(10,70){\line(1,0){20}}
\put(30,70){\line(0,-1){10}}
\put(30,60){\line(1,0){10}}
\put(40,60){\line(0,-1){10}}
\put(40,50){\line(1,0){10}}
\put(50,50){\line(0,-1){10}}

\put(-60,40){\mbox{$(R,P) \ \ =$}}

{\footnotesize
\put(123,17){\mbox{$ \bar P$}}
\put(17,50){\mbox{$R$}}
\put(243,22){\mbox{$\check P$}}
\qbezier(270,3)(280,20)(270,37)
\put(280,18){\mbox{$h_P = l_{P^{\vee}}=p_{_1}$}}
\qbezier(5,-5)(132,-20)(260,-5)
\put(130,-25){\mbox{$N $}}
\qbezier(5,35)(25,25)(45,35)
\put(22,20){\mbox{$l_R$}}
\qbezier(225,43)(245,52)(265,43)
\put(243,52){\mbox{$l_{\!_P}$}}
}

\put(4,40){\mbox{$\ldots$}}
\put(18,40){\mbox{$\ldots$}}
\put(32,40){\mbox{$\ldots$}}

\end{picture}

\noindent
where $l_{\!_P}$ is the number of lines in the Young diagram $P$.
This $(R,P)$ is
the first\footnote{``First" means here the representation that is associated with the Young diagram obtained just by attaching the diagram $R$ to $\bar P$ atop.} (hence, the word ``maximal") representation contributing to the product $R\otimes \bar P$. It can be manifestly obtained from the tensor products (i.e. as a projector from  $R\otimes \bar P$) by formula \cite{Koike}
\be
(R,P)=\sum_{Y,Y_1,Y_2}(-1)^{l_{\!_Y}}N^R_{YY_1}N^{P}_{Y^{\vee}Y_2}\ Y_1\otimes\overline{Y_2}
\ee
where $^\vee$ denotes the transposition of the Young diagram.

The adjoint representation in this notation is ${\rm adj} = ([1],[1])$,
while the conjugate of the representation $R$ is
$\overline{R} = (\varnothing,R)$.
The product
\be
[m]\otimes \overline{[m]} = \sum_{k=0}^m ([k],[k])
\ee
where $(\varnothing,\varnothing) \stackrel{SU(N)}{\cong} \varnothing$,
and the other items are diagrams with $2k$ lines, $k$ of length $N-1$ and $k$ of length one.
Similarly,
\be
[1^m]\otimes \overline{[1^m]} = \sum _{k=0}^m ([1^k],[1^k])
\ee
where contributing are the diagrams with just two lines of lengths $N-k$ and $k$.

The expression for the Hopf polynomial in arbitrary composite representations can be again written in the form
\be\label{30}
{\cal H}_{ (R,P)\times (T,S)} =
D_{(R,P)}\cdot \Sch_{(T,S)}\{p^{_*(R,P)}\}
\ee
Moreover, the main ingredients in this formula can be presented in a universal form that involves $N$ (if at all) as a simple parameter, however, the explicit expressions for them are far more involved:
$p^{_*(R,P)}_k$ implied by (\ref{plambdatimes}) for the composite Young diagram $(R,P)$ are \cite{ML,MMHopf}
\be
p_k^{*(R,P)} = q^{2{|R|-|P|\over N}k}\left(p_k^*+ \frac{1}{A^k}\cdot   \sum_{j=1}^{l_{\!_R}}  q^{(2j-1)k}\cdot(q^{-2kr_j}-1) + A^k\cdot  \sum_{i=1}^{l_P}q^{(1-2i)k}\cdot(q^{ 2kp_i}-1)
\right)
\label{compolocus}
\ee
the quantum dimensions are \cite{MMHopf}
\be\label{Dims}
D_{(R,P)}
=  D_{_R}(N-l_{\!_P})\, D_{_P}(N-l_{\!_R})\,
\frac{ \prod_{i=1}^{l_{\!_R}}[N-l_{\!P}-i]!\prod_{i'=1}^{l_{\!_P}}[N-l_{\!R}-i']!   }
{\prod_{i=1}^{l_{\!_R}+l_{\!_P} } [N-i]!} \, \prod_{i=1}^{l_{\!_R}}\prod_{i'=1}^{l_{\!_P}}
[N+r_i+p_{i'}+1-i-i']
\ee
where $[...]$ denotes quantum numbers, and the corresponding Schur functions are  \cite{Koike,Kanno,MMHopf}
\be
\Sch_{(R,P)}\{p^{*(T,S)}\}=
\sum_{\eta\in R\cap P^{\vee} } (-)^{|\eta|}\cdot\Sch_{R/\eta}\{p^{*(T,S)}\}
\cdot \Sch_{P/\eta^{\vee}}\{p^{*(T,S)}(A^{-1},q^{-1})\}
\label{compoSchur}
\ee
where $\Sch_{R/\eta}$ denotes the skew Schur function. Note that the ``mirror-reflecting" substitute
$(A,q)\to (A^{-1},q^{-1})$ in (\ref{compoSchur}) could be replaced just by the transposition of Young diagrams
except for the $U(1)$-factor $q^{2{|R|-|P|\over N}k}$ that have to be accordingly changed.

\section{Resolved conifold: Hopf link versus $L_{8n8}$}

Surprisingly or not, the topological vertex formalism provides expressions,
which are different from (\ref{L8n8}), but in an interesting way.
The four-leg diagrams from section \ref{coni}
can be also considered
as two glued 3-leg topological vertices with summation over intermediate
states.

The topological vertex
which is used in all-genus topological string calculations on toric Calabi-Yau 3-fold,
is given by  \cite{Aganagic:2003db, Okounkov:2003sp}\footnote{One can make the change $q\to 1/q$ so that the Schur polynomials would contain exponentials $q^{-\rho_0}$ like in formulas of sec.3. This would require the identification $Q=A^{-2}$ and reversing arrows in the figure below, and interchanging representations in the composite representation: $(R,P)\to (P,R)$.}
\beq
C_{\xi\mu\lambda} (q)  =
q^{\varkappa(\lambda)}\cdot \Sch_{\mu}( q^{\rho_0})
\sum_{\eta} \Sch_{\xi/\eta} (q^{\mu + \rho_0})\cdot  \Sch_{\lambda^\vee/\eta} (q^{\mu^\vee+ \rho_0}),
\label{TV}
\eeq
where $\varkappa(\lambda) = 2\sum_{i,j\in\lambda}(j-i)$, and the time variables are
\be
p_k^{(\mu)}  = p_k(q^{\mu+\rho_0}) = \sum_{j=1}^\infty  q^{(2\mu_j-2j+1)k} =
\frac{1}{q^{k}-q^{-k}} + \sum_{j=1}^\infty q^{(1-2j)k}(q^{2\mu_jk}-1)=
\frac{1}{q^{k}-q^{-k}}+(q^k-q^{-k})\sum_{i,j\in\mu} q^{2k(j-i)},
\label{pmu}
\ee
Note that they are independent of $A$ and different
from $p^{*\mu}_k$ in (\ref{plambdatimes}).
In topological string theory, the parameter $q$ is associated with the string coupling
(genus expansion parameter) $g_s$ by $q= e^{-g_s/2}$.
Since $\eta$ in the skew-characters should be a sub-diagram of $\lambda$, the sum over
$\eta$ is actually finite.
Though it is not manifest in \eqref{TV}, $C_{\xi^\vee\mu\lambda^\vee}(q) \cdot q^{-\varkappa(\mu)}$ is symmetric
under the cyclic permutation of $(\xi, \mu, \lambda)$.

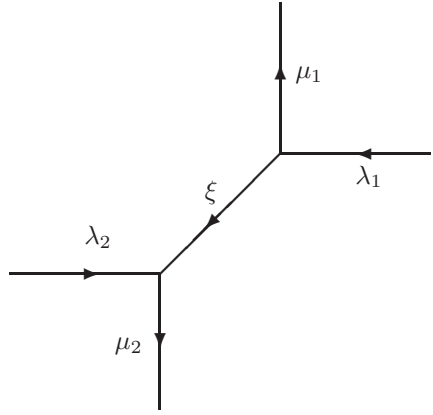
\unitlength 2mm
\begin{picture}(40,30)(-20,2)
\thicklines
\put(10,12){\vector(0,-1){5}}
\put(10,12){\line(-1,0){5}}
\put(10,12){\line(1,1){4}}
\put(10,3){\line(0,1){5}}
\put(0,12){\vector(1,0){6}}
\put(18,20) {\vector(-1,-1){5}}
\put(18,20){\line(1,0){5}}
\put(28,20){\vector(-1,0){5}}
\put(18,30){\line(0,-1){5}}
\put(18,20){\vector(0,1){6}}
\put(13,17){$\xi$}
\put(23,18){$\lambda_1$}
\put(5,14){$\lambda_2$}
\put(19,25){$\mu_1$}
\put(7,7){$\mu_2$}
\end{picture}
\unitlength 0.35mm

\noindent
\lq\lq Four point function\rq\rq\ on the resolved conifold geometry is given by
gluing two topological vertices:
\beq
Z_{\mu_1, \mu_2 ; \lambda_1, \lambda_2}
= \sum_\xi (-Q)^{\vert \xi \vert} C_{\xi \mu_1^\vee \lambda_1}(q)
C_{\xi^\vee \mu_2^\vee \lambda_2}(q),
\label{4pt}
\eeq
where $Q=  e^{t}$ and $t$ is the K\"ahler parameter of the rational curve $\mathbf{P}^1$
represented by the internal edge.
The prescription of large $N$ duality tells
that the 't Hooft coupling
is $t = N g_s = \frac{2\pi i N}{N+k}$, and we may identify $Q=q^{2N}=A^2$.
In the precise formulation, the diagram is equipped with arrows which
distinguish between the  representations and their transposed.
In writing down \eqref{4pt}, we set the clockwise order of edges at each vertex.
If the edge is outgoing, we have to use the transpose of the Young diagram attached to the edge.
To make the summation over $\xi$ easy,  we put $\xi$ to the first position
by using the cyclic symmetry of the topological
vertex. In the toric diagram, the slope of the edge corresponds to a cycle of $T^2$ fibration,
in particular,
the horizontal (vertical) line corresponds to $(1, 0)$($(0,1)$) cycle, respectively.
Hence we expect that
$\mu_{1,2}$ and $\lambda_{1,2}$ are linked with linking number $\pm 1$,
while $\mu_{1}(\lambda_1)$ and $\mu_{2}(\lambda_2)$ are parallel to each other.
Note that the notation is adjusted to purposes of the present paper,
where the cyclic ordering is less important and can be made obscure.

Our main claims in this paper are:
\begin{itemize}
\item
the normalized sum (\ref{4pt}) is equal to
\be\label{main}
\boxed{
\begin{array}{c}
\hat{\cal G}^{L_{8n8}}_{\mu_1\times\lambda_1\times\mu_2\times\lambda_2} =
\displaystyle{\frac{Z_{\mu_1,\mu_2;\lambda_1,\lambda_2}}
{Z_{\varnothing,\varnothing:\varnothing, \varnothing}}} =(-A)^{|\mu_1|+|\mu_2|}q^{\varkappa(\lambda_1)+
\varkappa(\lambda_2)+\varkappa(\mu_1)+\varkappa(\mu_2)}\cdot
D_{(\mu_1,\mu_2)}\times
\cr
\cr
\times\sum_{\sigma,\eta_1,\eta_2}(-A^2)^{|\eta_1|+|\eta_2|-|\sigma|}\cdot
\Sch_{\lambda_1^\vee/\eta_1}\{p^{(\mu_1^\vee)}\}\cdot\Sch_{\lambda_2^\vee/\eta_2}\{p^{(\mu_2^\vee)}\}
\cdot \Sch_{\eta_1^\vee/\sigma }\{p^{(\mu_2)}\}
\cdot\Sch_{\eta_2^\vee/\sigma^\vee}\{p^{(\mu_1)}\}
\end{array}
}
\ee
where the sums run over $\eta_1\subset\lambda_1^\vee$, $\eta_2\subset\lambda_2^\vee$ and $\sigma\subset\eta_1\cap\eta_2$. The quantum dimension of the composite representation, $D_{(\mu_1,\mu_2)}$ is given in (\ref{Dims}).
\item instead of the full unreduced HOMFLY of link $L_{8n8}$,
which in this case  is given by (\ref{L8n8}) and contains many terms:
\be
{\cal H}^{L_{8n8}}_{\lambda_1\times\mu_1\times\lambda_2\times\mu_2}
= \sum_{\stackrel{\lambda\in \lambda_1\otimes\bar\lambda_2}{\mu\in \mu_1\otimes\bar\mu_2}}
N^{\lambda}_{\lambda_1\lambda_2}\cdot N^\mu_{\mu_1\mu_2}\cdot
{\cal H}^{\rm Hopf}_{\lambda\times\mu}
\ee
the sum (\ref{main}) provides the contribution
of only the
maximal representations  $\lambda_{\rm max} \in \lambda_1\otimes\bar\lambda_2$
and $\mu_{\rm max} \in \mu_1\otimes\bar\mu_2$ so that
\be
\boxed{
{\cal G}^{L_{8n8}}_{\lambda_1\times\mu_1\times\lambda_2\times\mu_2}
= {\cal H}^{\rm Hopf}_{(\lambda_1,\lambda_2)\times(\mu_1,\mu_2)}=D_{(\mu_1,\mu_2)}\cdot \Sch_{(\lambda_1,\lambda_2)}\{p^{_*(\mu_1,\mu_2)}\}
}
\label{calG}
\ee
Here ${\cal G}$ differs from $\hat{\cal G}$ in (\ref{main}) by changing the framing to the standard one and taking into account the $U(1)$-factor:
\be\label{factors}
{\cal G}:=(-1)^{|\lambda_1|+|\lambda_2|+|\mu_1|+|\mu_2|}\cdot q^{-C_2(\lambda_1)-C_2(\lambda_2)-C_2(\mu_1)-C_2(\mu_2)}\cdot q^{2(|\lambda_1|-|\lambda_2|)(|\mu_1|-|\mu_2|)\over N}\cdot \hat {\cal G}
\ee
Here $C_2(R)=\varkappa(R)+N|R|$ is the eigenvalue of the second Casimir operator in the representation $R$, and we took into account that, for the composite representations, $C_2((R,S))=C_2(R)+C_2(S)$, \cite{GW}.

\end{itemize}
In the next two sections, we explain how to get formulas (\ref{main}), (\ref{calG}), and, in the Appendix, how the identity (\ref{calG}) works.

\section{Proof of (\ref{main})
\label{sumfor}}

Here we present summation formulas that lead to formula (\ref{main}).

We start with the simplest example: when a  diagram  at external line is empty, say, $\lambda =\varnothing$, then
the sum over $\eta$ in (\ref{TV}) is restricted to $\eta=\varnothing$
and
\be
C_{\xi\mu\varnothing} =\Sch_\mu(q^{\rho_0})\cdot \Sch_\xi(q^{\mu+\rho_0})
\label{lambda0first}
\ee
Then (\ref{4pt}) implies:
\be
Z_{\mu_1,\mu_2;\varnothing,\varnothing} \sim
\sum_\xi (- Q)^{|\xi|} \cdot\Sch_\xi(q^{\mu_1+\rho_0})
\cdot \Sch_{\xi^{\vee}}(q^{\mu_2 +\rho_0})
= \exp\left(-\sum_k   \frac{Q^k \,p_k^{(\mu_1)}p_k^{(\mu_2)}}{k}\right)
\ee
Substituting  (\ref{pmu}), we get for the $\mu$-dependent factor
\be
\exp\left(-\sum_k   \frac{Q^k \,p_k^{(\mu_1)}p_k^{(\mu_2)}}{k}\right) \sim
\exp \left\{-\sum_k   \frac{Q^k  }{k} \left(
\sum_{j}  \frac{q^{\mu^1_jk}-q^{-\mu^1_jk}}{q^k-q^{-k}}\cdot q^{(\mu^1_j-2j+1)k}
+\sum_j  \frac{q^{\mu_j^2k }-q^{-\mu^2_jk }}{q^k-q^{-k}}\cdot q^{(\mu^2_j-2j+1)k}
+\right.\right.\nn\\ \left.\left.
+ \sum_{j_1,j_2} (q^{2\mu^1_{j_1}k}-1)(q^{2\mu^2_{j_2}k}-1)q^{-2(j_1+j_2-1)k}
\right) \right\}
\ee
For particular diagrams $\mu_1$ and $\mu_2$, the ratios in this formula become polynomials,
and the entire expression is a product of factors $(1-Qq^{2n})$ with some $\mu$-dependent
powers $n$.
For example, if $\mu_1=\mu_2=[1]$, there are just two factors:
\be
\exp\left(-\sum_k   \frac{Q^k \,p_k^{([1])}p_k^{([1])}}{k}\right)
& \sim &
\exp\left\{-\sum_k \frac{Q^k}{k}
\Big(\overbrace{2+(q^{2k}-1)^2q^{-2k}}^{q^{2k}+q^{-2k}}\Big)\right\}
=\\
&=&(1-Qq^2)(1-Qq^{-2}) = A^2\Big(Aq-(Aq)^{-1}\Big)\Big(Aq^{-1}-A^{-1}q\Big)=A^2(q-q^{-1})^2D_{\rm adj}
\nonumber
\ee
Generally, restoring the $\mu$-independent factor
\be
\exp\left(-\sum_k   {1\over k}\cdot\frac{Q^k }{(q^k-q^{-k})^2}\right)=\prod_{i=1}^\infty (1-Qq^{-2i})^i
\ee
one obtains\footnote{Alternatively, one can derive this formula using instead of $p_k^{(\mu)}$ the components of $q^{\mu+\rho_0}=q^{2\mu_j-2j+1}$ in the Cartan plane, (\ref{pmu}). Then, the l.h.s. of (\ref{exp}) is a product
\be
\exp\left(-\sum_k   \frac{Q^k \,p_k^{(\mu_1)}p_k^{(\mu_2)}}{k}\right)=\prod_{i,j\ge 1}\Big(1-Qq^{2(\mu^1_i+\mu_j^2-i-j+1)}\Big)
\ee
which straightforwardly gives the r.h.s. of (\ref{exp}), see \cite{EC} for further details.}
\be\label{exp}
\exp\left(-\sum_k   \frac{Q^k \,p_k^{(\mu_1)}p_k^{(\mu_2)}}{k}\right) = (-A)^{|\mu_1|+|\mu_2|}
q^{\varkappa(\mu_1)+\varkappa(\mu_2)\over 2}h_{\mu_1}h_{\mu_2} D_{(\mu_1,\mu_2)}\prod_{i=1}^\infty (1-Qq^{-2i})^i
\ee
Here $h_\mu:=\prod_{i,j\in\mu}(q^{h_{i,j}}-q^{-h_{i,j}})$, $h_{i,j}$ is length of the hook $(i,j)$, and the quantum dimension $D_{(\mu_1,\mu_2)}$ is given in (\ref{Dims}).

Let us now note that
\be\label{spec0}
 \Sch_{\mu}( q^{\rho_0})={q^{\varkappa(\mu)\over 2}\over h_\mu}
 \ee
Hence, we finally obtain
 \be\label{35}
{ Z_{\mu_1,\mu_2;\varnothing,\varnothing}\over Z_{\varnothing,\varnothing;\varnothing,\varnothing}}= (-A)^{|\mu_1|+|\mu_2|}
q^{\varkappa(\mu_1)+\varkappa(\mu_2)}D_{(\mu_1,\mu_2)}
 \ee
which is a particular case of (\ref{main}) and, being proportional to the quantum dimension of $(\mu_1,\mu_2)$, is consistent with (\ref{calG}) at $\lambda_1=\lambda_2=\varnothing$.

\bigskip

When diagrams $\lambda$ are also non-trivial,  non-trivial can also be $\eta$ and
summation over the intermediate representation $\xi$
can be performed by the Cauchy formula for skew Schur functions
\be\label{sumC}
&&\sum_{\xi} (-Q)^{|\xi|} \cdot\Sch_{\xi/\eta_1}\{p\}\cdot\Sch_{\xi^\vee/\eta_2}\{p'\}=\\
&=&\exp\left(-\sum_k \frac{Q^kp_kp_k'}{k}\right) \cdot
\sum_\sigma (-Q)^{|\eta_1|+|\eta_2|-|\sigma|}\cdot
\Sch_{\eta_1^\vee /\sigma}\{p'\}\cdot\Sch_{\eta_2^\vee /\sigma^\vee}\{p\}
\nonumber
\ee
Now formulas (\ref{4pt}), (\ref{TV}), along with (\ref{exp}) and (\ref{sumC}), give rise to (\ref{main}).

\section{Proof of (\ref{calG})}

Now let us prove that the triple sum (\ref{calG}) reduces to the Hopf polynomial in the composite representation in the form (\ref{compoSchur}). To this end, we use the formula
\be\label{skew}
\Sch_{R/T}\{p^{(1)}+p^{(2)}\}=\sum_P\Sch_{R/P}\{p^{(1)}\}\cdot\Sch_{P/T}\{p^{(2)}\}
\ee
Let us start studying formula (\ref{main})
with the case of $\lambda_2 = \mu_2 =\varnothing$. In this case, the sum in (\ref{main}) reduces to
\be
\sum_{\eta}(-A^2)^{|\eta|}\cdot\Sch_{\lambda_1^\vee/\eta}\{p^{(\mu_1^\vee)}\}\cdot\Sch_{\eta^\vee}\{p^{(\varnothing)}\}&=&
\sum_{\eta}A^{2|\eta|}\cdot\Sch_{\lambda_1/\eta^\vee}\{-p^{(\mu_1^\vee)}\}\cdot\Sch_{\eta^\vee}\{p^{(\varnothing)}\}=\nonumber\\
=\sum_{\eta}A^{|\lambda_1|}\cdot\Sch_{\lambda_1/\eta^\vee}\{-A^{-k}p^{(\mu_1^\vee)}_k\}\cdot\Sch_{\eta^\vee}
\left\{{A^k\over q^k-q^{-k}}\right\}&=&A^{|\lambda_1|}\cdot\Sch_{\lambda_1}\left\{{A^k\over q^k-q^{-k}}-A^{-k}p^{(\mu_1^\vee)}\right\}=\nonumber\\
=A^{|\lambda_1|}\cdot\Sch_{\lambda_1}\left\{p^*_k + A^{-k}\sum_i q^{(2i-1)k}(q^{-2k\mu^1_i}-1)\right\}&=&
q^{-{2|\lambda_1||\mu_1|\over N}}A^{|\lambda_1|}\cdot\Sch_{\lambda_1}\left\{p^{*\mu_1}\right\}
\label{40}
\ee
where we used that
\be\label{id}
\Sch_{R^\vee/T^\vee}\{p_k\}=(-1)^{|R|+|T|}\Sch_{R/T}\{-p_k\}, \ \ \ \ \ \ p^{(\mu_1^\vee)}_k=(-1)^{k+1}p^{(\mu_1)}_k\Big|_{q\to -1/q}
\ee
and formula (\ref{skew}) with $T=\varnothing$.
Taking into account all the factors in (\ref{main}) and (\ref{factors}), we immediately obtain from (\ref{calG}) formula (\ref{10}) for the Hopf link.

\bigskip

In complete analogy, in the general case, using (\ref{skew}) we can evaluate in (\ref{main}) the sum over $\eta_1$
\be
&&\hspace{-1cm}\sum_{\eta_1}(-A^2)^{|\eta_1|}\cdot
\Sch_{\lambda_1^\vee/\eta_1}\{p^{(\mu_1^\vee)}\}\cdot \Sch_{\eta_1^\vee/\sigma }\{p^{(\mu_2)}\}=
A^{|\sigma|}(-A)^{|\lambda_1|}\cdot\sum_{\eta_1}
\Sch_{\lambda_1/\eta_1^\vee}\{-A^{-k}p^{(\mu_1^\vee)}_k\}\cdot \Sch_{\eta_1^\vee/\sigma }\{A^kp^{(\mu_2)}_k\}=\nonumber\\
&=&A^{|\sigma|}(-A)^{|\lambda_1|}\cdot
\Sch_{\lambda_1/\sigma}\{A^kp^{(\mu_2)}_k-A^{-k}p^{(\mu_1^\vee)}_k\}=
q^{-{2(|\lambda_1|-|\sigma|)(|\mu_1|-|\mu_2|)\over N}}A^{|\sigma|}(-A)^{|\lambda_1|}\cdot
\Sch_{\lambda_1/\sigma}\{p^{*(\mu_1,\mu_2)}\}
\ee
and, similarly, we evaluate the sum over $\eta_2$:
\be
\sum_{\eta_1}(-A^2)^{|\eta_2|}\cdot
\Sch_{\lambda_2^\vee/\eta_2}\{p^{(\mu_2^\vee)}\}\cdot \Sch_{\eta_2^\vee/\sigma^\vee }\{p^{(\mu_1)}\}=
q^{{2(|\lambda_2|-|\sigma|)(|\mu_1|-|\mu_2|)\over N}}A^{|\sigma|}(-A)^{|\lambda_2|}\cdot
\Sch_{\lambda_2/\sigma^\vee}\{p^{*(\mu_1,\mu_2)}(A^{-1},q^{-1})\}
\ee
Hence, we are remaining with
\be
(-A)^{|\lambda_1|+|\lambda_2|}q^{{2(|\lambda_2|-|\lambda_1|)(|\mu_1|-|\mu_2|)\over N}}\sum_\sigma (-1)^{|\sigma|}\cdot\Sch_{\lambda_1/\sigma}\{p^{*(\mu_1,\mu_2)}\}\cdot
\Sch_{\lambda_2/\sigma^\vee}\{p^{*(\mu_1,\mu_2)}(A^{-1},q^{-1})\}
\ee
which is exactly the sum in (\ref{compoSchur}). Now, taking into account all the factors in (\ref{main}), (\ref{factors}), we obtain (\ref{calG}).

\section{Symmetry properties of Hopf polynomials}

The main tools of Hopf calculus in \cite{tangcalc} are the recursion (\ref{HvsHH}),
which is a characteristic feature of characters and thus
implies (\ref{HopfthroughSchur}),
and a peculiar property of the Hopf link,
\be
{\cal H}^{\rm Hopf}_{R_1\times\bar R_2}(A,q) = {\cal H}^{\rm Hopf}_{R_1\times R_2}(A^{-1},q^{-1})
\label{conjHopf}
\ee
which is obvious from the picture

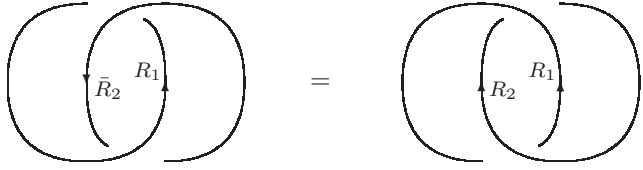
\begin{picture}(300,100)(-120,-50)

\qbezier(-30,0)(-30,30)(0,30)
\qbezier(-30,0)(-30,-30)(0,-30)
\qbezier(30,0)(30,-30)(0,-30)

\qbezier(30,0)(30,20)(22,24)

\put(30,0){
\qbezier(-30,0)(-30,30)(0,30)
\qbezier(30,0)(30,30)(0,30)
\qbezier(30,0)(30,-30)(0,-30)

\qbezier(-30,0)(-30,-20)(-22,-24)
}

\put(0,1){\vector(0,-1){2}}
\put(30,-1){\vector(0,1){2}}
{\footnotesize
\put(3,-5){\mbox{$\bar R_2$}}
\put(18,2){\mbox{$R_1$}}
}

\put(85,-2){\mbox{$=$}}

\put(150,0){
\qbezier(-30,0)(-30,30)(0,30)
\qbezier(-30,0)(-30,-30)(0,-30)
\qbezier(30,0)(30,30)(0,30)

\qbezier(30,0)(30,-20)(22,-24)

\put(30,0){
\qbezier(-30,0)(-30,-30)(0,-30)
\qbezier(30,0)(30,30)(0,30)
\qbezier(30,0)(30,-30)(0,-30)

\qbezier(-30,0)(-30,20)(-22,24)
}

\put(0,-1){\vector(0,1){2}}
\put(30,-1){\vector(0,1){2}}

{\footnotesize
\put(3,-5){\mbox{$R_2$}}
\put(18,2){\mbox{$R_1$}}
}

}

\end{picture}

\noindent
supplemented by the property of ${\cal R}$-matrix, that its inversion is equivalent to
inversion of $A$ and $q$.
Note that the
equality (\ref{conjHopf}) specifically holds for the Hopf link despite ${\cal R}_{R_1\times\bar R_2}(A,q)\neq {\cal R}_{R_1\times R_2}(A^{-1},q^{-1})$.

Other symmetry properties for the Hopf polynomials in the case of composite representations are
\be\label{Id1}
{\cal H}^{\rm Hopf}_{(R,P)\times S} =
{\cal H}^{\rm Hopf}_{ (P,R)\times S}(A^{-1},q^{-1})
\ee
which follows from the relation $\overline{(R,P)}=(P,R)$, and
\be\label{LRD}
{\cal H}^{\rm Hopf}_{(R,P)\times (S,T)} =(-1)^{|R|+|P|+|S|+|T|}q^{4(|R|-|P|)(|S|-|T|)\over N}
{\cal H}^{\rm Hopf}_{ (R^\vee,P^\vee)\times (S^\vee,T^\vee)}(A,q^{-1})
\ee
which is a Hopf version of the standard level-rank duality, \cite{LRD}
\be
H_{\{R_i\}}=H_{\{R_i^\vee\}}(A,q^{-1})\nonumber
\ee
the latter valid for any HOMFLY invariant being a fact from group representation theory. Formula (\ref{LRD}) follows from (\ref{30}) and (\ref{compoSchur}) using (\ref{compolocus}) and (\ref{id}).

From the three identities (\ref{conjHopf})-(\ref{LRD}), it follows that
\be
{\cal H}^{\rm Hopf}_{(R,P)\times [1]} =(-1)^{|P|+|R|+1}q^{4(|P|-|R|)\over N}
{\cal H}^{\rm Hopf}_{ (P^\vee,R^\vee)\times [1]}(A^{-1},q)
\label{RPPR}
\ee
In fact, the change of variable $A \to A^{-1}$ is natural from the viewpoint of conifold geometry.
Under the flop operator of conifold, which flips the sign of the K\"ahler parameter and hence $A \to A^{-1}$,
the representation attached with $\lambda_1=[1]$ changes from $\mu_1$ to $\mu_2$. Hence, this identity.

While formulas of the type (\ref{HopfthroughSchur}) and their generalizations to the composite representations \cite{Kanno,MMHopf} as well as the Rosso-Jones formula are convenient for general studies of the Hopf link, the most efficient for obtaining general explicit formulas is at the moment the tangle calculus of \cite{tangcalc}. Making use of these identities and one additional calculation for
${\cal H}^{\rm Hopf}_{{\rm adj}\times [s]}$,
the old knowledge of Hopf polynomials in two symmetric representations
${\cal H}^{\rm Hopf}_{[r]\times[s]}$ can be extended to the case of one arbitrary diagram and one symmetric.

Explicit formulas for the Hopf polynomials that we use in the Appendix are all picked up from \cite{tangcalc}.

\section{Rosso-Jones formula and superpolynomials}

There is another, basically very different formula for the Hopf polynomial based on the paper by M. Rosso and V. F. R. Jones \cite{RJ}
\be
{\cal H}_{\lambda\times \mu}^{\rm Hopf}=q^{2|\lambda||\mu|\over N}
q^{\varkappa_\lambda+\varkappa_\mu} \sum_{\eta\in \lambda\otimes \mu}
N_{\lambda\mu}^\eta \cdot q^{-\varkappa_{\eta}}\cdot D_{\eta}
\label{RJ}
\ee
In this paper, we explained how the representation of the Hopf polynomial as a specialization of the character, (\ref{HopfthroughSchur}) is directly related with the topological vertices. However, any direct relation either of these two to the Rosso-Jones formula (\ref{RJ}) is not clear, to our best knowledge. All known proofs of equivalencies with this relation are provided by an additional averaging procedure, either in Chern-Simons theory, \cite{Lab}, or via (related) a multiple integral representation \cite{BEMT}, or using a scalar product of two characters defined in some other way \cite{EK}. There is, certainly, a derivation based on the modular categories \cite{MK}, i.e. basically on the Verlinde formula \cite{Ver}, however, it is not straightforward as well.
As a particular manifestation of this problem, generalizing the Rosso-Jones formula for other knots and links, one can not obtain a counterpart of the character specialization (see, however, \cite{HL}).

Thus, we have a diagram of the type

\bigskip

$$
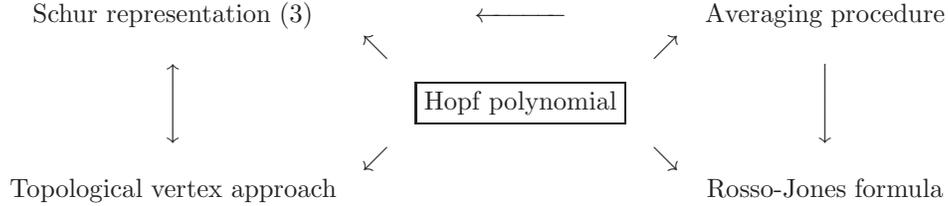
\begin{array}{ccccc}
\hbox{Schur representation  (\ref{HopfthroughSchur})}&&\xleftarrow{\hspace*{1cm}}&&\hbox{Averaging procedure}\\
&\nwarrow&&\nearrow &\\
\Bigg\updownarrow&&\boxed{\hbox{Hopf polynomial}}&&\Bigg\downarrow\\
&\swarrow&&\searrow&\\
\hbox{Topological vertex approach}&&&&\hbox{Rosso-Jones formula}
\end{array}
$$

\bigskip

\noindent
It would be interesting to restore the missing arrows.

What is, however, important, the Rosso-Jones formula admits very immediate and simple generalization to any other torus knots and links \cite{RJ}, and also to the superpolynomials \cite{DMMSS}, while other ingredients at the diagram are only easily generalizable to the superpolynomials \cite{EK,AS,IK}. Indeed, the Rosso-Jones formula for the Hopf superpolynomial looks like (hereafter in this section, we omit the $U(1)$-factor)
\be
{\cal P}_{\lambda,\mu}^{\rm Hopf} =
q^{-\nu_\lambda-\nu_\mu}t^{\nu'_{\lambda}+\nu'_{\mu}} \sum_{\eta\in \lambda\otimes \mu}
\mathfrak{N}_{\lambda\mu}^\eta \cdot q^{\nu_\eta}t^{-\nu'_{\eta}}
\cdot {\cal M}_\eta
\label{supRJHopf}
\ee
where $\nu_\lambda:=2\sum_i (i-1)\lambda_i$, $\nu'_\lambda:=\nu_{\lambda^\vee}$ so that $\varkappa_\lambda=\nu'_\lambda-\nu_\lambda$, and $M_\eta$ is the Macdonald dimension of $\eta$, i.e. the specialization of the Macdonald symmetric function $M_\eta \{q,t|p\}$ at the topological locus (in time variables) $p_k=p_k^*$ \cite{DMMSS}:
\be
{\cal M}_\eta:=M_\eta \{q,t|\mathfrak{p}^*\},\ \ \ \ \ \ \ \mathfrak{p}_k^*={A^k-A^{-k}\over t^k-t^{-k}}
\ee
where we use the Gothic letters in order to stress the superpolynomial deformation. In particular, the coefficients $\mathfrak{N}_{\lambda\mu}^\eta$ are defined now as coefficients of expansion of the products of Macdonald polynomials
\be
M_\lambda \{q,t|p\}\cdot M_\mu \{q,t|p\}=\sum_{\eta\in \lambda\otimes \mu}\mathfrak{N}_{\lambda\mu}^\eta\cdot M_\eta \{q,t|p\}
\ee
In contrast with the Littlewood-Richardson coefficients, $\mathfrak{N}_{\lambda\mu}^\eta$ are not obligatory integer.

A counterpart of formula (\ref{HopfthroughSchur}) in the superpolynomial form looks as
\be
{\cal P}_{\lambda,\mu}^{\rm Hopf} =
M_\lambda(t^{-\rho})\cdot M_\mu(q^{-\lambda}t^{-\rho})={\cal M}_\lambda\cdot M_\mu(q^{-\lambda}t^{-\rho})
\label{supthroughSchur1}
\ee
These symmetric functions of the components of vectors in the Cartan plane can be again rewritten in terms of the time variables
\be
\mathfrak{p}_k^{*\lambda} = \mathfrak{p}^*_k - A^{-k}(q^k-q^{-k})\sum_{i,j\in\lambda}t^{k(2i-1)}q^{k(1-2j)}=\mathfrak{p}^*_k + A^{-k}\sum_i t^{(2i-1)k}(q^{-2k\lambda_i}-1)
\ee
where, as above, the Gothic letter refers to the superpolynomial deformation:
\be
M_\mu(q^{-\lambda}t^{-\rho})=M_\mu \{q,t|\mathfrak{p}_k^{*\lambda}\}
\ee
Thus, (\ref{supthroughSchur1}) can be written in the form
\be\label{supHopf}
{\cal P}_{\lambda,\mu}^{\rm Hopf} ={\cal M}_\lambda\cdot M_\mu \{q,t|\mathfrak{p}_k^{*\lambda}\}
\ee
The equivalence of the two representations for the Hopf superpolynomials, (\ref{supRJHopf}) and (\ref{supHopf}) is again proved through an intermediate averaging representation \cite{EK,AS,IK}.

Now one can repeat the machinery developed in the present paper, in this superpolynomial case, \cite{AKMMM}.

\section{Conclusion}

In this letter, we considered an elementary example of the tangle calculus of \cite{tangcalc}:
the quadratic recursion formula (\ref{HvsHH}) for Hopf polynomials,
which immediately follows from  pictorial gluing of free ends of the Hopf tangle.
We explained the relation to the traditional description of Hopf polynomials,
which provides an algebraic proof  of the recursion identity from multiplication of
Schur functions.
This proof is easily lifted to superpolynomials, and emerging identity
suggests that tangle calculus can be extended to this area,
despite the Reshetikhin-Turaev formalism, of which it is a simple corollary,
is no longer applicable in this case in any known way.
This adds a new evidence to the comparably surprising results of \cite{DMMSS}
and \cite{AnoMevoKR}
about survival of evolution property for the torus superpolynomials and
even for the torus Khovanov-Rozansky polynomials at finite $N$.
As another application, we described the relation of the link polynomial for
$L_{8n8}$ and of the Hopf polynomial.

To summarize this application, the link $L_{8n8}$,
which one could naturally associate with the 4-point toric diagram,
in knot theory is distinguished by existence of two dual descriptions:
as a closed necklace made from 4 unknots and as a Hopf link in reducible
representations $ (\mu_1\otimes\mu_2)\times(\lambda_1\otimes\lambda_2)$.

\begin{picture}(300,180)(-200,-135)

\qbezier(-40,0)(-40,20)(0,20) \qbezier(40,0)(40,20)(0,20)
\qbezier(-40,-10)(-40,-30)(0,-30) \qbezier(40,-10)(40,-30)(0,-30)

\put(0,-80){
\qbezier(-40,0)(-40,20)(0,20) \qbezier(40,0)(40,20)(0,20)
\qbezier(-40,-10)(-40,-30)(0,-30) \qbezier(40,-10)(40,-30)(0,-30)
}

\qbezier(-40,-5)(-60,-5)(-60,-45)\qbezier(-40,-85)(-60,-85)(-60,-45)
\qbezier(-40,-5)(-20,-5)(-20,-25)\qbezier(-40,-85)(-20,-85)(-20,-65)
\qbezier(-19.5,-32)(-19,-45)(-19.5,-58)

\qbezier(40,-5)(60,-5)(60,-45)\qbezier(40,-85)(60,-85)(60,-45)
\qbezier(40,-5)(20,-5)(20,-25)\qbezier(40,-85)(20,-85)(20,-65)
\qbezier(19.5,-32)(19,-45)(19.5,-58)

\put(-60,-45){\vector(0,1){2}}
\put(-19,-45){\vector(0,-1){2}}
\put(0,20){\vector(1,0){2}}
\put(0,-30){\vector(-1,0){2}}

\put(60,-45){\vector(0,-1){2}}
\put(19,-45){\vector(0,1){2}}
\put(0,-110){\vector(-1,0){2}}
\put(0,-60){\vector(1,0){2}}

\put(-75,-20){\mbox{$\lambda_2$}}
\put(-20,25){\mbox{$\mu_1$}}
\put(65,-20){\mbox{$\lambda_1$}}
\put(-20,-120){\mbox{$\mu_2$}}

\end{picture}

\noindent
The possibility of  two different descriptions allows one to double-check the formulas,
see \cite{tangcalc}, but here we  used only the Hopf-related expressions.
Our claim (\ref{calG}) was that the sum (\ref{4pt}) provides the contribution of the
composite representations
\be
{\cal G}^{L_{8n8}}_{\lambda_1\times\mu_1\times\lambda_2\times\mu_2}
= {\cal H}^{\rm Hopf}_{(\mu_1,\mu_2)\times (\lambda_1,\lambda_2)}
\label{calG1}
\ee
rather than the full unreduced HOMFLY polynomial.

\bigskip

It is an interesting task to extend all the components of our discussion:
tangle calculus, character calculus, conifold calculus and their refinement
beyond the Hopf link example.
Some parts of this extension are already long-standing problems,
however, the interplay between these different approaches seems to provide
new tools to finally solve them and make new steps  towards
building a powerful and efficient calculational technique.

\section*{Acknowledgements}

Our work is supported in part by Grants-in-Aid for Scientific Research
(\# 17K05275) (H.A.), (\# 15H05738) (H.K.) and JSPS Bilateral Joint Projects (JSPS-RFBR collaboration)
``Topological Field Theories and String Theory: from Topological Recursion
to Quantum Toroidal Algebra'' from MEXT, Japan. It is also partly supported by the grant of the Foundation for the Advancement of Theoretical Physics ``BASIS" (A.Mor.), by  RFBR grants 16-01-00291 (A.Mir.) and 16-02-01021 (A.Mor.), by joint grants 17-51-50051-YaF, 18-51-05015-Arm (A.M.'s).

\section*{Appendix. Explicit examples of formula (\ref{calG})}

Here we give a series of examples that illustrate how formula (\ref{calG}) works. First of all, we will need a series of specializations of the Schur functions, some of them being an illustration of formula (\ref{spec0}):
\beqa
\Sch_{[1]} (q^{\rho_0}) &=& p_1(q^{\rho_0}) = (q - q^{-1} )^{-1}, \CR
\Sch_{[2]} (q^{\rho_0}) &=& \frac{1}{2} \Big((p_1(q^{\rho_0}))^2 + p_2(q^{\rho_0})\Big) =
(q - q^{-1} )^{-2}(1 + q^{-2} )^{-1},  \CR
\Sch_{[1^2]} (q^{\rho_0}) &=& \frac{1}{2} \Big((p_1(q^{\rho_0}))^2 - p_2(q^{\rho_0})\Big)
= (q - q^{-1} )^{-2}(1 + q^2 )^{-1}.
\eeqa
Other answers that we will need in examples are
\beqa
\Sch_{[1]} (q^{[1] + \rho_0}) &=& (q - q^{-1} ) + \Sch_{[1]} (q^{\rho_0}),
\CR
\Sch_{[1]} (q^{[2] + \rho_0}) &=&  (1+ q^2) (q - q^{-1} ) + \Sch_{[1]} (q^{\rho_0}),
\CR
\Sch_{[1]} (q^{[1^2] + \rho_0}) &=& (1+ q^{-2}) (q - q^{-1} )  + \Sch_{[1]} (q^{\rho_0}),
\label{2box}
\eeqa

\subsection*{\underline{$\lambda_1 = \lambda_2 =\varnothing$}}

This case was already considered in (\ref{35}). Using formula (\ref{Dims}), we can compare
\beqa
E([1],[1]) &\sim &
\exp\left\{-\sum_k \frac{Q^k}{k}
\Big(q^{2k}+q^{-2k}\Big)\right\}
=(1-Qq^2)(1-Qq^{-2})=A^2(q-q^{-1})^2[N+1][N-1]\nonumber\\
\hbox{with}\ \ \ \ {\cal G}^{L_{8n8}}_{[1] \times\varnothing\times [1] \times\varnothing}
&=& [N+1][N-1] =  D_{{\rm adj}}
= {\cal H}^{\rm Hopf}_{([1],[1])\times\varnothing}, \CR
E([2],[1])
&\sim &
\exp\left\{-\sum_k \frac{Q^k}{k}
\Big(q^{4k}+1+q^{-2k}\Big)\right\}
=(1-Qq^4)(1-Q)(1-Qq^{-2})=\nonumber\\
&=&A^3(q-q^{-1})^3[N+2][N][N-1]\nonumber\\
\hbox{with}\ \ \ \ {\cal G}^{L_{8n8}}_{[2] \times\varnothing\times [1] \times\varnothing}
&=&\frac{1}{[2]} [N+2][N][N-1] =D_{([2],[1])}
= {\cal H}^{\rm Hopf}_{([2],[1])\times\varnothing}, \CR
E([1,1],[1])
&\sim&
\exp\left\{-\sum_k \frac{Q^k}{k}
\Big(q^{2k}+1+q^{-4k}\Big)\right\}
=(1-Qq^2)(1-Q)(1-Qq^{-4})=\nonumber\\
&=&A^3(q-q^{-1})^3[N+1][N][N-2]\nonumber\\
\hbox{with}\ \ \ \ {\cal G}^{L_{8n8}}_{[1,1] \times\varnothing\times [1] \times\varnothing}
&=& \frac{1}{[2]} [N+1][N][N-2] = D_{([1,1],[1])}
= {\cal H}^{\rm Hopf}_{([1,1],[1])\times\varnothing}, \CR
E([2],[2])
&\sim&
\exp\left\{-\sum_k \frac{Q^k}{k}
\Big(q^{6k}+2+q^{-2k}\Big)\right\}
=(1-Qq^6)(1-Q)^2(1-Qq^{-2})=\nonumber\\
&=&A^4(q-q^{-1})^4[N+3][N]^2[N-1]\nonumber\\
\hbox{with}\ \ \ \ {\cal G}^{L_{8n8}}_{[2] \times\varnothing\times [2] \times\varnothing}
&=& \frac{1}{[2]^2} [N+3] [N]^2 [N-1] = D_{([2],[2])}
= {\cal H}^{\rm Hopf}_{([2],[2])\times\varnothing}, \CR
E([1,1],[1,1])
&\sim&
\exp\left\{-\sum_k \frac{Q^k}{k}
\Big(q^{2k}+2+q^{-6k}\Big)\right\}
=(1-Qq^2)(1-Q)^2(1-Qq^{-6})=\nonumber\\
&=&A^4(q-q^{-1})^4[N+1][N]^2[N-3]\nonumber\\
\hbox{with}\ \ \ \ {\cal G}^{L_{8n8}}_{[1,1] \times\varnothing\times [1,1] \times\varnothing}
&=& \frac{1}{[2]^2} [N+1] [N]^2 [N-3] = D_{([1,1],[1,1])}
= {\cal H}^{\rm Hopf}_{([1,1],[1,1])\times\varnothing}, \CR
E([2],[1,1])
&\sim&
\exp\left\{-\sum_k \frac{Q^k}{k}
\Big(q^{4k}+q^{2k}+q^{-2k}+q^{-4k}\Big)\right\}
=(1-Qq^4)(1-Qq^2)(1-Qq^{-2})(1-Qq^{-4})=\nonumber\\
&=&A^4(q-q^{-1})^4[N+2][N+1][N-1][N-2]\nonumber\\
\hbox{with}\ \ \ \ {\cal G}^{L_{8n8}}_{[2] \times\varnothing\times [1,1] \times\varnothing}
&=& \frac{1}{[2]^2}[N+2] [N+1] [N-1] [N-2] = D_{([2],[1,1])}
= {\cal H}^{\rm Hopf}_{([2],[1,1])\times\varnothing},
\eeqa
where we used the notation $\exp\left(-\sum_k   \frac{Q^k \,p_k^{(\mu)}p_k^{(\nu)}}{k}\right):=E(\mu,\nu)$.

\subsection*{\underline{$ \lambda_1 = [1], \lambda_2 =\varnothing$}}

In this case we have
\beq
Z_{\mu_1, \mu_2 ; [1] ,\varnothing}
= \Sch_{\mu_1} (q^{\rho_0})~\Sch_{\mu_2} (q^{\rho_0})
\sum_\xi (-Q)^{\vert \xi \vert}
\sum_{\tau} \Sch_{\xi^\vee/\tau}( q^{\mu_1 + \rho_0} )~\Sch_{[1]/\tau}( q^{\mu_1^\vee+ \rho_0} )
~\Sch_\xi( q^{\mu_2 + \rho_0} ),
\eeq
where we can write down the summation over $\tau$ explicitly;
\beq
\sum_{\tau} \Sch_{\xi^\vee/\tau}( q^{\mu_1 + \rho_0} )~\Sch_{[1]/\tau}( q^{\mu_1^\vee+ \rho_0} )
= \Sch_{\xi^\vee}( q^{\mu_1 + \rho_0} )~\Sch_{[1]}( q^{\mu_1^\vee+ \rho_0} )  + \Sch_{\xi^\vee/[1] }( q^{\mu_1 + \rho_0} ).
\label{decomp}
\eeq
Performing the summation over $\xi$ by the Cauchy formula, we obtain
\beq
Z_{\mu_1, \mu_2 ; [1] ,\varnothing}
= Z_{\mu_1, \mu_2 ; \varnothing ,\varnothing}
\left( \Sch_{[1]} ( q^{\mu_1^\vee + \rho_0}) - Q \cdot \Sch_{[1]} (q^{\mu_2 + \rho_0}) \right).
\eeq
Now, using (\ref{2box}), we obtain
\beqa
{\cal G}^{L_{8n8}}_{[1] \times [1] \times [1] \times\varnothing} (A,q)
&= &[N+1][N] [N-1] (q^2 -1 + q^{-2})
= {\cal H}^{\rm Hopf}_{([1],[1])\times [1]} \CR
{\cal G}^{L_{8n8}}_{[2] \times [1] \times [1] \times\varnothing} (A,q)
&=& q^{{2/ N}}\cdot\frac{ [N+2][N][N-1]}{[2]}
\Big( [3][N] - A^{-1} q^{-2} (q  - q^{-1})^2 \Big)
 = {\cal H}^{\rm Hopf}_{([2],[1])\times [1]}
 \CR
{\cal G}^{L_{8n8}}_{[2] \times [1] \times [2] \times\varnothing} (A,q)
&=& \frac{[N+3] [N]^2 [N-1]}{[2]^2}
\Big( [3][N]+ [N+2] (q - q^{-1})^2 \Big)
= {\cal H}^{\rm Hopf}_{([2],[2])\times [1]}    \CR
{\cal G}^{L_{8n8}}_{[2] \times [1] \times [1,1] \times\varnothing} (A,q)
&=& \frac{[N+2] [N+1][N] [N-1] [N-2]}{[2]^2} (q^4 - 1 + q^{-2})
= {\cal H}^{\rm Hopf}_{([2],[1,1])\times [1]} \nonumber
\eeqa
and all other cases of the Young diagrams up to the level 2 are obtained from these using formulas (\ref{Id1})-(\ref{RPPR}).

\subsection*{\underline{$ \lambda_1 = \lambda_2 = [1]$}}

In this case we have
\beqa
Z_{\mu_1, \mu_2 ; [1], [1] }
&=& \Sch_{\mu_1} (q^{\rho_0})~\Sch_{\mu_2} (q^\rho_0)
\sum_\xi (-Q)^{\vert \xi \vert}
\sum_{\tau} \Sch_{\xi^\vee/\tau}( q^{\mu_1 + \rho_0} )~\Sch_{[1]/\tau}( q^{\mu_1^\vee+ \rho_0} ) \CR
&&~~ \times
\sum_{\sigma} \Sch_{\xi /\sigma}( q^{\mu_2 + \rho_0} )~\Sch_{[1]/\sigma}( q^{\mu_2^\vee+ \rho_0} ).
\eeqa
Using \eqref{decomp}, we find
\beq
 {\cal G}^{L_{8n8}}_{\mu_1 \times [1] \times \mu_2 \times [1]}
=  {\cal G}^{L_{8n8}}_{\mu_1 \times \varnothing \times \mu_2 \times \varnothing } \cdot z(\mu_1, \mu_2),
\eeq
where
\beqa
z(\mu_1, \mu_2) &:=&
\Sch_{[1]} (q^{\mu_1^\vee + \rho_0})~\Sch_{[1]} (q^{\mu_2^\vee + \rho_0})
 -Q (1 + \Sch_{[1]} (q^{\mu_1 + \rho_0})~\Sch_{[1]} (q^{\mu_1^\vee + \rho_0})  \CR
&&~~ +  \Sch_{[1]} (q^{\mu_2 + \rho_0})~\Sch_{[1]} (q^{\mu_2^\vee + \rho_0}) )
+ Q^2 \cdot \Sch_{[1]} (q^{\mu_1 + \rho_0})~\Sch_{[1]} (q^{\mu_2 + \rho_0})),
\eeqa
which is symmetric under $\mu_1 \leftrightarrow \mu_2$.
Thus we obtain
\beq
{\cal G}^{L_{8n8}}_{[1] \times [1] \times [1] \times [1]}
&=&  [N+1][N-1] \Big( -1 + [3]^2[N]^2 \Big)
 = {\cal H}^{\rm Hopf}_{{\rm adj}\times {\rm adj}}
\CR
{\cal G}^{L_{8n8}}_{[2] \times [1] \times [1] \times [1]}
&=& {[3]\over [2]}\ [N+2][N+1][N][N-1]\Big((q^3+q^{-3})[N]-[N+1]\Big) = {\cal H}^{\rm Hopf}_{([2],[1])\times {\rm adj}},
\CR
{\cal G}^{L_{8n8}}_{[2] \times [1] \times [2] \times [1]}
&=&  \frac{[N+3][N+1][N]^2[N-1]}{[2]^2}\Big((q^3+q^{-3})^2[N+1]-2(q^3+q^{-3})[N+2]+[N+3]\Big) =  {\cal H}^{\rm Hopf}_{([2],[2])\times {\rm adj}}, \CR
{\cal G}^{L_{8n8}}_{[2] \times [1] \times [1,1] \times [1]}
&=&  \frac{[N+2][N+1][N-1][N-2]}{[2]^2}\Big((q^3+q^{-3})[N]-[N+1]\Big)\Big((q^3+q^{-3})[N]-[N-1]\Big)={\cal H}^{\rm Hopf}_{([2],[1,1])\times {\rm adj}}\CR
\eeqa

\subsection*{\underline{$ \lambda_1 = [2]~\mathrm{or}~[1,1] , \lambda_2 =\varnothing$}}

For $\lambda_2 = [2]$, we have
\beq
Z_{\mu_1, \mu_2 ; [1] ,\varnothing}
= \Sch_{\mu_1} (q^{\rho_0})~\Sch_{\mu_2} (q^\rho_0) \sum_\xi (-Q)^{\vert \xi \vert}
\sum_{\tau} \Sch_{\xi^\vee/\tau}( q^{\mu_1 + \rho_0} )~\Sch_{[2]/\tau}( q^{\mu_1^\vee+ \rho_0} )
~\Sch_\xi( q^{\mu_2 + \rho_0} ),
\eeq
where
\beqa
&&\sum_{\tau} \Sch_{\xi^\vee/\tau}( q^{\mu_1 + \rho_0} )~\Sch_{[2]/\tau}( q^{\mu_1^\vee+ \rho_0} ) \CR
&&~~= \Sch_{\xi^\vee}( q^{\mu_1 + \rho_0} )~\Sch_{[2]}( q^{\mu_1^\vee+ \rho_0} )
+ \Sch_{\xi^\vee/[1]}( q^{\mu_1 + \rho_0} )~\Sch_{[1]}( q^{\mu_1^\vee+ \rho_0} )
+ \Sch_{\xi^\vee/[2] }( q^{\mu_1 + \rho_0} ).
\label{decomp2}
\eeqa
The Cauchy formula gives
\beq
 {\cal G}^{L_{8n8}}_{\mu_1 \times [2] \times \mu_2 \times \varnothing}
=  {\cal G}^{L_{8n8}}_{\mu_1 \times \varnothing \times \mu_2 \times \varnothing}
\left( \Sch_{[2]} ( q^{\mu_1^\vee + \rho_0}) - Q \cdot \Sch_{[1]} (q^{\mu_1^\vee + \rho_0})~\Sch_{[1]} (q^{\mu_2 + \rho_0})
+ Q^2 \cdot \Sch_{[1,1]} (q^{\mu_2 + \rho_0})\right).
\eeq
When $\mu_1 = \mu_2 = [1]$, we obtain
\beqa
{\cal G}^{L_{8n8}}_{[1] \times [2] \times [1]  \times \varnothing}
&=& A^{2} [N+1][N-1] \left( s_{[2]} ( q^{[1] + \rho_0})
- Q s_{[1]} (q^{[1] + \rho_0})^2 + Q^2 s_{[1,1]} ( q^{[1] + \rho_0}) ) \right) = \CR
 &=&\frac{[N+1][N][N-1]}{[2]}\Big((q^3+q^{-3})[N+2]-[N+3]\Big)
= {\cal H}^{\rm Hopf}_{{\rm adj}\times [2]}
\eeqa

On the other hand, if $\lambda_1 = [1,1]$, $\Sch_{[2]}$ and
$\Sch_{[1,1]}$ switch places everywhere, and the parallel computation  gives
\beq
A^4 \cdot {\cal G}^{L_{8n8}}_{[1] \times [1,1] \times [1]  \times \varnothing} (A, q)
= A^4 \cdot {\cal G}^{L_{8n8}}_{[1] \times [2] \times [1]  \times \varnothing}(A, q^{-1})
= {\cal H}^{\rm Hopf}_{{\rm adj}\times [1,1]}.
\eeq


\end{document}